# USING A MONTE-CARLO-BASED APPROACH TO EVALUATE THE UNCERTAINTY ON FRINGE PROJECTION TECHNIQUE

Jerome Molimard and Laurent Navarro
LCG, UMR 5146,
École Nationale Supérieure des Mines, CIS-EMSE, CNRS, Saint-Étienne, France
molimard@emse.fr, navarro@emse.fr

**ABSTRACT:**
A complete uncertainty analysis on a given fringe projection set-up has been performed using Monte-Carlo approach. In particular the calibration procedure is taken into account. Two applications are given: at a macroscopic scale, phase noise is predominant whilst at microscopic scale, both phase noise and calibration errors are important. Finally, uncertainty found at macroscopic scale is close to some experimental tests (~100 µm).

## 1. INTRODUCTION

The confidence on the result obtained by Optical full field techniques (OFFT) is poorly described, and the error estimation is millstone in their diffusion in industrial world. Usually, the measuring chain is complex, implying optical elements, numerical processing (correlation, phase extraction ...) and post-processing (derivation, filtering …). A lot of work have been carried out in order to improve and/or characterize each element of the measuring chain, in particular for image correlation [1] [2] or phase extraction [3]. Again, some experimental work gives a global sight on errors, see for example [4]. Anyway, overall measurement error still never have been achieved, in particular because of the difficulties to integrate different error sources, among them errors due calibration procedure. Prediction through error model is complex and usually cannot be achieved using standard error propagation rules. Previous works show the efficiency of Monte-Carlo based procedure on specific element of the measuring chain: description of the error on phase extraction [5]; post-processing derivation [6], optimal position of illumination vectors in 3D ESPI [7]. Anyway, no global prediction approach has been carried out in so far authors know.
The present work describes a generic way to estimate overall error in fringe projection [8], either due to random sources or the bias (calibration errors). Here, a high level calibration procedure based on pinhole model has been implemented [9]. The Monte Carlo procedure requires complete models of the calibration procedure and of the reference experiment. Here, the reference experiment consists in multiple steps out-of-plane displacement of a plane surface. First results at a macroscopic level are given, and a comparison with experimental ground truth made.

## 2. FRINGE PROJECTION

### 2.1. Description of the fringe projection set-up

The optical set-up for 3D measurement is a classical fringe projection set-up, with a pocket-projector 3M MPRO 110 of 1200×800 pixel resolution and an 8 bits CCD camera Imaging Source of 1280×960 pixels resolution. This solution is adapted to fields of investigation from 10×7 mm$^2$ to 200×150 mm$^2$ (see figure 1). The system uses a low-distortion lens (Linos, 0.3× f/8), inducing a very low error (less than 10$^{-4}$).

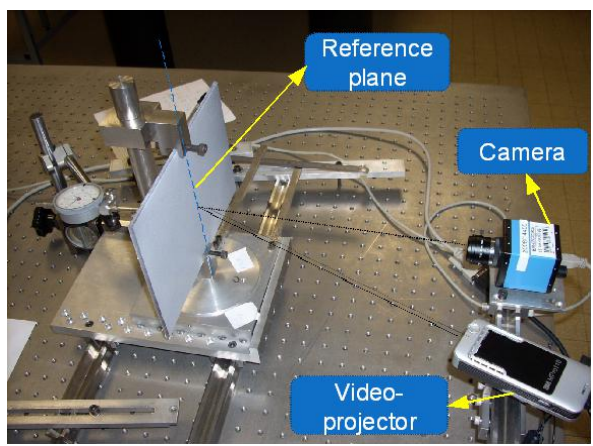

**Figure 1. Optical set-up and calibration test-rig.**

### 2.2. Implementation of the fringe projection

The physical principle is straightforward: a periodic pattern is projected on an object; the light is diffused by the object and captured by a CCD video-camera. The deformation of the fringes, recorded as phase maps, has a known dependency to the out-of-plane displacements of the illuminated object. Light intensities on an object illuminated by a set of fringes can be described by a periodic function with a perturbation $\phi = \phi(x,y)$ corresponding to the object shape, as expressed by:

$$\phi(x,y) = \frac{2\pi \times \tan\theta(x,y)}{p(x,y)} z(x,y) \qquad (1)$$

In this expression, the sensitivity characterized by the slope of the linear relationship between Φ(x,y) and z(x,y), can be adjusted by modifying the pitch *p* or the angle θ between the CCD camera and the video-projector. It has to be noted that the sensitivity varies locally, due to the divergent optical arrangement of the video projector and the CCD camera. Here, the pin-hole model is chosen to describe this situation because it is simple and therefore open to interpretation. Parameters of the model are:
- the camera magnification along the vertical axis ($\gamma_{CCD}$) and along the horizontal axis $\gamma_{CCD}/\tau_{CCD}$,
- the distance between the CCD camera and the reference plane ($h_0$),
- the distance between the video-projector and the reference plane ($h_p$),
- the distance between the video-projector focal point and the CCD camera axis (*d*).

Now, the point A(x, y, z) is known for any position in the object plane, referred by the co-ordinates M(r, s) applying the pin-hole model:

$$z(r,s) = \frac{h_p h_o \left[ (2\pi f_p h_p - P_t d \varphi) \frac{\gamma_{CCD}}{\tau_{CCD}} \times r - P_t (d^2 + h_p^2) \varphi(r,s) \right]}{h_0 \left[ (2\pi f_p h_p - P_t d \varphi) d + P_t (d^2 + h_p^2) \varphi(r,s) \right] - h_p \left[ 2\pi f_p h_p - P_t d \varphi(r,s) \right] \frac{\gamma_{CCD}}{\tau_{CCD}} \times r}$$

$$x(r,s) = \frac{z(r,s) + h_0}{h_0} \frac{\gamma_{CCD}}{\tau_{CCD}} \times r \qquad (2)$$

$$y(r,s) = \frac{z(r,s) + h_0}{h_0} \gamma_{CCD} \times s$$

Last, the phase is obtained through temporal phase shifting and discrete Fourier Transform:

$$\varphi(r,s) = \arctan_{2\pi} \left( \frac{\sum_{k=1}^{nq} \left\{ \sin\left(k \cdot \frac{2\pi}{q}\right) I_k(r,s) \right\}}{\sum_{k=1}^{nq} \left\{ \cos\left(k \cdot \frac{2\pi}{q}\right) I_k(r,s) \right\}} \right) \qquad (3)$$

### 2.3. Calibration procedure

Measuring all the parameters of a the pin-hole model is difficult in practice and an inverse calibration is more adapted. Here, the calibration is based on the known rotation of a reference plane [9]. The calibration procedure is in two steps: first, the phase map of a plane perpendicular to the camera axis is taken. Second, the plan is rotated along the vertical axis, and second phase map is recorded. Even if the method is straightforward, some hypothesis should be fulfilled: video-projector and camera axis should converge on a single point, this point being on the rotation axis; rotation axis should be perpendicular to the plane defined by the camera axis and the video-projector axis, and parallel to the fringes.
A complete experimental strategy has been established to fulfil these requirements; after setting up the fringe projection arrangement, the quality can be checked by analysing the shape of each calibration plane: it is possible to have an experimental estimate of the plane tilting, and rotation axis verticality and centring; the calibration is validated if the tilting of the plane or the rotation axis is lower than 1/10th of millimetre, this value being a minimum adjustable value considering the set-up. Last, the phase quality can be estimated by comparing the theoretical phase surface to the experimental one. After convenient correction if necessary, the error distribution shows that the phase error can be modelled as a random Gaussian distribution at a global level.
Note that the system has to be calibrated after each geometrical changes in the configuration, but not before each new experiment.

### 2.4. Example of experimental result

A reference experimental test has been designed: it consists in a sphere cut in a plate. The system has a standard macroscopic design, with no special optimization for this particular specimen. Result is presented Figure 2.
The depression in the plate has been estimated using least square approximations of a sphere and of a plate. The deepness is defined by the length between the lower sphere point and its projection on the plate. Because of averaging effects due to the high number of measurement points, this value becomes almost insensitive to random noise. Measured value is 2.10 mm and the sphere radius to 15.17 mm. High frequency component of the experimental field is used to estimate random phase noise. Its standard deviation is representative of the common values encountered in the laboratory (σ = 43 μm). The results obtained for the sphere deepness using a dial indicator is 1.93±0.03 μm. Difference between fringe projection and the dial indicator is 107 µm. This value can be related to some systematic errors, in particular due to the calibration procedure.

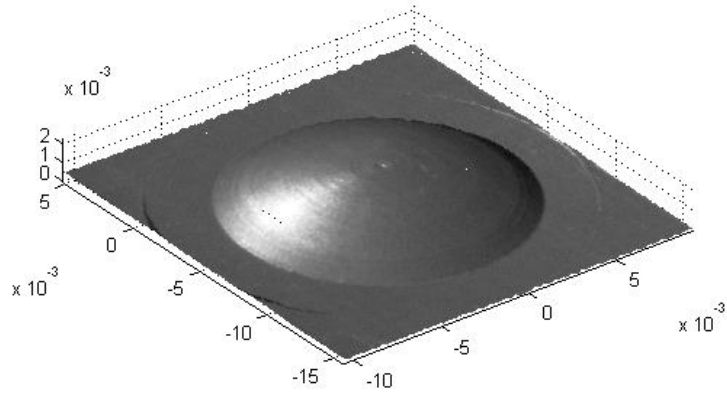

Figure 2. Reconstruction of a reference sphere-in-plate

## 3. ERROR EVALUATION

### 3.1. Measurement model

Within the Monte-Carlo framework, it is necessary to model the optical system, including its possible defects, and to give a probability density function (PDF) for each error source. In order to achieve such a goal, the approach will take into account the whole calibration procedure, giving an estimate of the calibration parameters and, later, on the $z(x, y)$ function. Calibration procedure used in this document is based on strong assumptions. Of course, in real conditions, these assumptions are not completely true, and errors should be taken into account in order to evaluate the global uncertainty level on the fringe projection process. The measuring system model implemented for this Monte-Carlo approach will consider the following uncertainties: the position of the reference plane, an error on apparent pixel size, and an error on the rotation value. Other error sources, summarized as a random error on the phase measurement, are related the intensity measurement and the phase extraction [5]. Last, camera lens distortion might be added to the model; here anyway, this aspect is not taken into account because the model should become too complex compared to the influence of lens used in common experiments on out-of-plane information ($z$).

In order to have a good comparison on the different situations, some experimental data are necessary. A reference situation, corresponding to the laboratory practice is defined. The field of view is 68×54 mm2 and the sensitivity set to 5 mm per fringe. Resolution is supposed to be 1/100$^{th}$ fringe, i.e. 50 µm. Corresponding geometrical data are given in Table 1; the input positioning error is set 2 mm for each geometrical parameter, corresponding to a relative error around 1% (see Table 1). All calibration data – including PDFs – are summarized in Table 2. The values are evaluated from laboratory experience.

|  | $P_t/f_p$ (m/m) | $h_p$ (mm) | $h_0$ (mm) | $d$ (mm) |
|---|---|---|---|---|
| Reference value | 5.1 10$^{-3}$ | -474 | -340 | 341 |
| Uncertainty | 7 10$^{-4}$ | 2 | 2 | 2 |
| Error type | B | B | B | B |

Table 1. Geometrical parameters and uncertainties (macroscopic scale)

| Parameters | | PDF | Nominal value | ½ length or standard deviation | Error type |
|---|---|---|---|---|---|
| Calibration parameters | | | | | |
| | $\alpha$ | uniform | 2.1 ° to 4.2 ° | 0.14 ° | B |
| | $\gamma_{CCD}$ | gaussian | 53 µm | 0.06 µm | B |
| | $\tau_{CCD}$ | gaussian | 1 | 0.017 | B |
| Reference plane mispositionning | | | | | |
| | $\beta_1, \beta_2, \beta_3$ | gaussian | 0 | 3 ' | B |
| | $x_1, x_2, x_3$ | gaussian | 0 | 0.25 mm | B |
| Phase dispersion | | | | | |
| | $\delta\varphi$ | gaussian | 0 | $10^{-2} \times 2\pi$ | A |

Table 2. Calibration parameters and uncertainties (macroscopic scale)

## 3.2. Shape uncertainty

Study will be held 1/ considering independently each error source 2/ using independent sources all together. Each case study uses 40 random samples, giving a compromise between calculation time and precision. In order to evaluate the influence of calibration procedure on shape reconstruction, a very simple test is proposed: it consists in reconstructing the $z = 0$ mm plane, translating it and reconstructing it at the position $z = 1$ mm and $z = 2$ mm. The exact shape is completely known, the shape variation as well.

Analysis of the results is based first on the bias i.e. the mismatch between the mean and the expected value. Results show that the mean position is very close to the theoretical one on the reference in any conditions: bias error is found to be close to $2 \cdot 10^{-7}$ mm. The values estimated on the $z = 1$ mm and $z = 2$ mm planes are the same. Surprisingly, the global error level seems to be independent to the rotation angle. The standard deviation is considerably higher ($3 \cdot 10^{-4}$ m). The analysis of reconstructed maps clearly shows that the standard deviation amplitude on shape maps is mainly due to a deterministic effect: the position of the plane is rotated in space, or, in other words, the position of the virtual reference plane is erroneous. This problem should be considered in many cases as a minor problem.

Now, it is interesting to quantify how the calibration errors may induce a reconstruction error independently from the reference plane absolute position. The reconstruction error is divided into two contributions: a high frequency one, representing a random error, mainly related to the phase error, and a low-frequency one, related to calibration uncertainties. Even if the calibration uncertainties result in an erroneous curvature, in the following, they will only be characterized by a standard deviation. Calibration uncertainties represents 15 µm and the random noise 50 µm in the studied case (P = 66 %). As a consequence, the calibration seems sufficiently efficient, and efforts have to be put on the random phase noise. The total error on the instrument is in this situation 104 µm (P = 95 %), i.e. 2% of the measuring range, considering as a basic definition the height variation for one fringe step.

## 4. CONCLUSIONS

Error estimation on optical full field techniques (OFFT) is millstone in the diffusion of OFFT. The present work describes a generic way to estimate overall error in fringe projection, either due to random sources (phase error, basically related to the quality of the camera and of the fringe extraction algorithm) or the bias (calibration errors). Here, a high level calibration procedure based on pinhole model has been implemented. This model compensates for the divergence effects of both the video-projector and the camera.

The work is based on a Monte Carlo procedure. So far, complete models of the calibration procedure and of a reference experiment are necessary. Here, the reference experiment consists in multiple step out-of-plane displacement of a plane surface. Using this very simple test, it is possible to observe that:

1- the uncertainties in the calibration procedure lead to a global rotation of the plane ; this means that a surface is reconstructed in a frame of reference slightly different from the global frame of reference of the experimental set-up. As a matter of fact, a variation between a reference position and a stressed one becomes independent from this parameter.
2- the overall error for a classical macroscopic set-up, with standard noise level, is 104 µm.
3- the main error source is the phase error ($\sigma = 50$ µm).
4- the calculated values are very reasonable compared with experimental ground truth ($\sigma = 43$ µm, bias ≈ 107 µm).

Finally, the aim of such a tool is to give some quantitative data on the overall uncertainty; this work can be easily used to determine before experiments the performance of a fringe projection set-up. In the near future, this approach will be used to optimize a microscopic, high sensitivity fringe projection set-up.